\documentstyle[12pt]{article}

\begin{document}

\begin{center}
{\Large \bf Testing Deca-TeV Unified Compositeness}\\
{\Large \bf at the 4 TeV $\bf\mu^+\mu^-$  Collider}\footnote{Presented 
by Yu.~F.~Pirogov at the
International Workshop on Linac-Ring Tipe $e\,p$ and $\gamma\,p$ Colliders, 
9-11 April 1997, Ankara.}\\

\vspace{4mm}

Vasily V.~Kabachenko 
 and  Yury F.~Pirogov\footnote{E-mail: pirogov@mx.ihep.su}\\[1ex]
{\it Theory Department, Institute for High Energy Physics,\\
Protvino, Moscow Region,  Russia\\
and\\
Moscow Institute of Physics and Technology,\\
Dolgoprudny, Moscow Region, Russia}

\end{center}

\begin{abstract}
\noindent
In the framework of the unified compositeness of leptons, quarks and 
Higgs bosons, the hidden local symmetry $\hat H_{loc}=SU(2)_L\times U(1)_Y$ 
with the heavy composite vector bosons, in addition
to the SM gauge bosons, is briefly described. Supplementary hypothesis of the
vector boson dominance (VBD) of the SM gauge interactions is 
considered. It is argued that this should produce the universal dominant
residual interactions of the SM composite particles, i.e.,  all of the fermions 
and Higgs bosons.  Restrictions on 
the universal residual fermion-fermion, fermion-boson and boson-boson
interactions due to the VBD are investigated. Manifestations of the residual
interactions at the 4~TeV $\mu^+\mu^-$   collider are studied. It is 
shown 
that at 95\%~C.L.~the unified substructure could be investigated at the 
collider 
in the processes $\mu^+\mu^-\to \bar ff$ up to
the compositeness scale
${\cal O}(150\,\,$TeV), in the processes $\mu^+\mu^-\to ZH$, $W^+W^-$
up to ${\cal O}
(100\,\,$TeV) and in the process $\mu^+\mu^-\to ZHH$ up to ${\cal O}(40\,\,
$TeV), which lie in the
naturally preferable Deca-TeV region. 
\end{abstract}

\section*{Introduction}
The scheme of the unified
compositeness of leptons, quarks and Higgs bosons, with constituents in common,
provides one of the promising ways to 
go beyond the Standard Model (SM) (for a short review, see~\cite{unified}).
Treating the SM Higgs doublet as  Goldstone boson in the scheme,
 one can solve, in particular,
 the  naturalness problem  of the Higgs sector in the SM without supersymmetry.
A nonlinear model was investigated in the 
lines described above by one of the present authors (Yu.F.P.) in refs.~\cite
{pirogov1,pirogov2}. Here
the SM is to be considered just a renormalizable part of the ``low energy'' 
effective field theory caused by the unified compositeness.

The effective ``low energy'' theory of the unified compositeness is
based on some rather general assumptions about symmetry properties.
Let the hypothetical hyperstrong interactions responsible for the internal 
binding of the SM composite
particles posses a global chiral symmetry $G$. Under the 
hyperstrong confinement,
 the symmetry
$G$ breakes down to some of its subgroup $H\subset G$ at the scale ${\cal F}$.
 In this, the
true Goldstone bosons appear which are ultimately identified, in particular, 
as the Higgs doublet. The unbroken symmetry $H$ must
contain the SM symmetry $SU(2)_L\times U(1)_Y$. 
Thus at the first stage, the
electroweak symmetry remains unbroken. Ultimate taking into account the 
gauge quantum corrections, corresponding to some extended electroweak
symmetry $I_{loc}\subset G$, results in the SM electroweak symmetry breaking at 
the Fermi scale 
$v\ll{\cal F}$. If this breaking happens only under 
two-loop  corrections, the naturalness relation between the scales $v$ 
and ${\cal F}$
takes place: ${\cal F}={\cal O}(2m_W/\alpha_W)$. So ${\cal F}$ is expected
 to lie naturally in the Deca-TeV region: ${\cal F}={\cal O}(10\,\,$TeV).
The minimal extension of the SM symmetry to implement such a scenario is given
by the choice  $G=SU(3)_L\times U(1)$ and $H=SU(2)_L\times
U(1)_Y$, the intrinsic local subgroup being $I_{loc}=SU(2)_L\times
U(1)_Y\times U(1)_{Y'}$. The corresponding nonlinear model $G/H$ may be
called the Minimal Nonlinear Standard Model (MNSM).

In what follows, we describe in short the linearization of the model via the
phenomenon of the hidden local symmetry. Then we present the crucial 
phenomenological consequences of the unified compositeness at the future 
4 TeV ${\mu^+\mu^-}$  collider (see, e.~g., refs.~\cite{mu94}--\cite{bar97}).

\section{Universal Residual Interactions}
As the nonlinear model, the MNSM is built on the nonlinear realization of $G$ 
that becomes linear when restricted to $H$~\cite{coleman}. Such a model  is
equivalent,
at least at the classical level,
 to the model with linearly realized symmetry $G\times
\hat H_{loc}$~\cite{kugo}. Here $\hat H_{loc}$ is the hidden local 
symmetry with the appropriate auxiliary
gauge bosons. In the context of the MNSM the phenomenon of the hidden local 
symmetry was studied in ref.~\cite{pirogov2}. The essence of the latter one
is as 
follows.

In the linear model, the field variable is the element of the whole 
group $G$ which 
can be parametrized as:
\begin{equation}
\hat \xi=\xi h,\,\,\,h\in H
\end{equation}
and
\begin{equation}
\xi=e^{i\phi'Y'/{\cal F}'}e^{i(\phi_\alpha X^{\dagger\alpha}+h.c.)/{\cal F}}\in 
G/H.
\end{equation}
Here $\phi$ is the Higgs-Goldstone doublet, $\phi'$ is the Goldstone boson
corresponding to the broken hypercharge  $Y'$, with ${\cal F}$ and ${\cal F}'$ 
being the symmetry breaking mass scales.
The following transformation law under $\gamma\times\hat h(x)\in G\times 
\hat H_{loc}$ takes 
place:
\begin{equation}
\gamma\times \hat h(x):\,\,\hat\xi\to \gamma\hat \xi\hat h^\dagger(x).
\end{equation}
The linear model describes spontaneous/dynamical  symmetry breaking
$G\times \hat H_{loc}\,\to\,H$, with the total local symmetry being broken as 
$I_{loc}\times\hat H_{loc}
\,\to\,H_{loc}=SU(2)_L\times U(1)_Y$.

To construct the Lagrangian of the linear model one has to introduce
the modified differential 1-form $\hat\omega_\mu=1/i\,\hat \xi^\dagger 
\hat D_\mu 
\hat \xi$, with
 $\hat D_\mu$ being the derivative covariant both under the intrinsic 
gauge symmetry $I_{loc}$ and the hidden local symmetry $\hat H_{loc}$.
Let us divide $\hat\omega_\mu$ into two parts: $\hat\omega_{\|\mu}$ which is
 parallel to $G/H$  and $\hat\omega_{\perp\mu}$ orthogonal
to it.
Under 
$G\times
\hat H_{loc}$ the parallel part $\hat\omega_{\|\mu}$ transforms homogeneously 
as in the original nonlinear model,
 and so does now the orthogonal part $\hat\omega_{\perp\mu}$. 
It is precisely introducing 
the auxiliary vector fields $\hat W_\mu^i$ and 
$\hat S_\mu$, corresponding to $\hat H_{loc}$,
that makes the
transformation of $\hat\omega_\perp$ homogeneous.  In the unitary under
$\hat H_{loc}$ gauge, i.e., at $h\equiv 1$ in Eq.~1, the modified 1-form 
looks like 
\begin{eqnarray}
\hat \omega_{\|\mu}&=&\omega_{\|\mu},\nonumber\\
\hat \omega_{\perp\mu}^i&=& \omega^i_{\perp\mu}-\hat g\hat W^i_\mu,\\
\hat \omega^0_{\perp\mu}&=&\omega^0_{\perp\mu}-\hat g_1\hat S_\mu,
\nonumber
\end{eqnarray}
where $\omega_\mu$ is the 1-form present in the original MNSM,
 $\hat g$ and $\hat g_1$ being some new strong coupling constants (expectedly,
$\hat g^2/4\pi ={\cal O}(1)$ and similarly for $\hat g_1$).
 
In the Lagrangian of the linear model, the new terms appear.
 They are related with the orthogonal part of the modified 1-form.
Here are some of the appropriate terms in the gauge sector:
\begin{equation}
\frac{\lambda{\cal F}^2}{2}(\hat\omega_{\perp\mu}^i)^2+\frac{\lambda_1
{\cal F}^2}{2}(\hat\omega_{\perp\mu}^0)^2+\cdots,
\end{equation}
and for fermions they are
\begin{eqnarray}
&&\bar\psi\gamma_\mu i(\partial_\mu+i\hat g\hat W^i_\mu T^i+i\hat g_1\hat S_\mu
)\psi\nonumber\\
&&+\kappa\bar\psi\gamma_\mu T^i\psi\hat\omega_{\perp\mu}^i+\kappa_1\bar\psi
\gamma_\mu Y\psi\hat\omega_{\perp\mu}^0+\cdots.
\end{eqnarray}
Here $\lambda$'s and $\kappa$'s are free parameters. It's to be noted that the
matter fields $\psi$ transform now only under $\hat H_{loc}$. The modified
covariant derivative  for them contains only the composite
$\hat W_\mu$ and $\hat S_\mu$, but not the elementary $W_\mu$ and $S_\mu$,
the latter ones entering only through the nonminimal interactions.

Introducing the vector fields in such a way  without kinetic terms 
is just a formal procedure. But we believe that the required 
kinetic terms are developed
by the quantum effects, and the new composite vector bosons become physical. 
This takes place, e.g., in 2- and 3-dimensional nonlinear 
$\sigma$-models~\cite{dada},
as well as in the hadron physics as accomplished fact.

From the Lagrangian of the linear model, one can read off the Lagrangian 
terms of the vector boson-current interactions:
\begin{eqnarray}
{\cal L}_{int}&=&-gW_\mu^i\Bigl((1-\lambda)J_\mu^i(\phi)
+\kappa J_\mu^i(\psi)\Bigr)
\nonumber\\
&&-\hat g \hat W_\mu^i\Bigl(\lambda J_\mu^i(\phi)+(1-\kappa)J_\mu^i(\psi)
\Bigr).  
\end{eqnarray}
Here $J_\mu^i(\psi)=\bar\psi\gamma_\mu T^i\psi$ and $J_\mu^i(\phi)=
\phi^\dagger i\tau^i/2\!\stackrel{\leftrightarrow}{D}_\mu\!\phi$ are the 
usual SM isotriplet currents, with $D_\mu$ being the
SM covariant derivative. To these isospin terms, one has to add the similar 
hypercharge isosinglet terms.
Impose now the natural requirement that all the composite particles $\phi$ and 
$\psi$ 
interact directly only with the composite vector bosons $\hat W$ and $\hat S$,
but not with the elementary ones $W$ and $S$. In  other words, this is
 the well-known
hypothesis of the vector boson dominance (VBD).
This requirement allows one to fix the free parameters: $\lambda=1$,
$\kappa=0$ and similarly for the isosinglet parameters.

The terms $(\hat\omega^i_\perp)^2$ and $(\hat\omega^0_\perp)^2$ 
describe the mass mixing of the elementary and composite gauge bosons,
namely,
$W$ with $\hat W$ and $S$ with $\hat S$. Diagonalizing these terms one gets 
 two sets of physical
vector bosons: the massless isotriplet and isosinglet physical 
bosons $\bar W^i$ 
and $\bar S$, as well as the massive ones $\bar{\hat W}^i$ and $\bar {\hat S}$
with masses of order  ${\cal F}$.
Due to the  heavy physical vector boson exchange, the new low energy effective
current-current interactions appear in addition to that of the SM:
\begin{eqnarray}
{\cal L}_{int}^{(VBD)}&=&-\frac{1}{2{\cal F}^2}\Bigl(J^i_\mu(\psi)J^i_\mu(\psi)
+\eta_1J^0_\mu(\psi)J^0_\mu(\psi)\Bigr)\nonumber\\
&&-\frac{1}{{\cal F}^2}\Bigl(J^i_\mu(\psi)J^i_\mu(\phi)+\eta_1J^0_\mu(\psi)
J^0_\mu(\phi)\Bigr).
\end{eqnarray}
Here $\eta_1$ is a free parameter, related to the original MNSM.
Note that the VBD does not affect the low energy Higgs boson 
self-interactions, the
latter ones being 
determined by the original MNSM alone:
\begin{eqnarray}
{\cal L}_{int}(\phi)=-\frac{1}{{\cal F}^2}\Bigl(\frac{1}{3}J^i_\mu(\phi)
J^i_\mu(\phi)+J^0_\mu(\phi)J^0_\mu(\phi)\Bigr)
\end{eqnarray}
(up to the Fiertz rearrangement).
All these expressions are valid only at energies $\sqrt{s}\ll{\cal F}$.

To resume, the unified compositeness plus the VBD prescribe the 
two-parameter set of the universal residual 
fermion-fermion, fermion-boson and boson-boson  interactions, 
with their space-time
and internal structure being fixed, sign including. The unified 
compositeness scale ${\cal F}$ is expected to lie in the Deca-TeV region.
Hence the TeV energies are required to probe these new contact interactions.

\section{Manifestations of VBD at ${\bf \mu^+\mu^-}$  Collider}
In a series of papers
we have investigated the possibility  to test the hypothesis of the VBD of
electroweak interaction at the future 2~TeV $e^+e^-$  linear collider via the
processes $e^+e^-\to \bar ff$, where $f=e^-,\mu^-,\tau^-,u,d,s,c,
b$~\cite{kabach1}, $e^+e^-\to ZH$, 
$W^+W^-$~\cite{kabach2} and $ZHH$~\cite{guryev}\footnote{This process was 
investigated with the CompHEP package for the symbolical and numerical 
calculations in the high energy physics \cite{comphep}.}. 
In this report
we have reconsidered the results for the future $\mu^+\mu^-$  collider
with the  total energy 4~TeV and the integrated luminocity 
10$^3\,\,fb^{-1}$~\cite{atac}, and found that this collider 
could present the definite answer about the existence (or opposite) of 
the Deca-TeV unified compositeness. 

To illustrate the dependence of the
observables on the parameters $\eta_1$ and ${\cal F}$, we present in what
follow the simple approximate formulas for differential cross-sections
for some of the processes.

$\mu^+\mu^-\to e^+e^-(\tau^+\tau^-)$:
\begin{eqnarray}
\frac{d\sigma(\mu^-_L)}{d\cos\theta}&=&\frac{\pi\bar\alpha^2}{4s}\biggl(
\kappa^2_1\frac{1}{16\bar c^4\bar s^4}(1+\cos\theta)^2+\kappa_2^2\frac{1}
{4\bar c^4}(1-\cos\theta)^2\biggr),\nonumber \\
\frac{d\sigma(\mu^-_R)}{d\cos\theta}&=&\frac{\pi\bar\alpha^2}{4s}\kappa_2^2
\biggl(\frac{1}{\bar c^4}(1+\cos\theta)^2+\frac{1}{4\bar c^4}
(1-\cos\theta)^2\biggr), 
\end{eqnarray}
here and in what follows $\bar c\equiv\cos\bar\theta_W$, $\bar s\equiv\sin\bar
\theta_W$ and $\bar\theta_W$ is the effective weak mixing angle and
 $\bar\alpha$
is the effective fine structure constant at energies under consideration. 
Scattering angle $\theta$ is that 
between $e^-$  and $\mu^-$.

$\mu^+\mu^-\to W^+W^-$:
\begin{eqnarray}
\frac{d\sigma(\mu^-_L)}{d\cos\theta}&=&\frac{\pi\bar\alpha^2}{4s}\biggl(
\kappa^2_1\frac{1}{16\bar c^4\bar s^4}+\frac{1}{4\bar s^4}\frac{u^2+t^2}{t^2}
\biggr)(1-\cos^2\theta),\nonumber \\
\frac{d\sigma(\mu^-_R)}{d\cos\theta}&=&\frac{\pi\bar\alpha^2}{4s}\kappa_2^2
\frac{1}{4\bar c^4}(1-\cos^2\theta),
\end{eqnarray}
here $\theta$ is the scattering angle between $W^-$ and $\mu^-$ and $s$, $t$, 
$u$ are the usual invariant kinematical variables.

$\mu^+\mu^-\to ZH$:
\begin{eqnarray}
\frac{d\sigma(\mu^-_L)}{d\cos\theta}&=&\frac{\pi\bar\alpha^2}{2s}\frac{E_Z}
{\sqrt s}\kappa_3^2\frac{(\bar s^2-\bar c^2)^2}{4\bar s^4\bar c^4}\frac
{E_Z^2}{s}(1-\cos^2\theta),\nonumber \\
\frac{d\sigma(\mu^-_R)}{d\cos\theta}&=&\frac{\pi\bar\alpha^2}{2s}\frac{E_Z}
{\sqrt s}\kappa_2^2\frac{1}{\bar c^4}\frac
{E_Z^2}{s}(1-\cos^2\theta),
\end{eqnarray}
$\theta$ is the scattering angle between $Z$ and $\mu^-$ and $E_Z$ is the c.m.\
$Z$ boson energy. 

The structure of these expressions is rather simple, namely, the appropriate
 SM contributions to  the
cross-sections are rescaled by factors
\begin{eqnarray}
\kappa_1&=&1-(1+\eta_1)\frac{\bar c^2\bar s^2}{\bar e^2}\frac{s}{{\cal F}^2},
\nonumber \\
\kappa_2&=&1-\eta_1\frac{\bar c^2}{\bar e^2}\frac{s}{{\cal F}^2},\\
\kappa_3&=&1-(\eta_1-1)\frac{\bar s^2\bar c^2}{\bar e^2(\bar s^2-\bar c^2)}
\frac{s}{{\cal F}^2}.
\end{eqnarray}
All these expressions are valid in the kinematical region $m_W^2,m_Z^2\ll
s,|t|\ll {\cal F}^2$ and are obtained in the high energy limit by
 neglecting the 
terms  ${\cal O}(m^2/{\cal F}^2,m^2/s)$ relative to these 
${\cal O}(s/{\cal F}^2)$\footnote{The net effect of the $\mu\mu H$
coupling ($\sim m_\mu$) in the total cross-sections of the 
processes $\mu^+\mu^-\to ZH$ and $ZHH$ 
proved to be numerically negligible at 
the energy under consideration.}. 
All the leading terms in this limit come from the
Lagrangian of Eq.~8, i.e., from the VBD interactions. Note that the cross-sections
for all the processes $\mu^+\mu^-\to\bar ff$ (except for  $\mu^+\mu^-\to
\mu^+\mu^-$) have the same structure as that for $\mu^+\mu^-\to e^+e^-$ (Eq.~10)
with the same rescaling factors. Similarly for the process $\mu^+\mu^-\to
ZHH$ relative to that  $\mu^+\mu^-\to ZH$ , but here $(1-\cos^2\theta)$ in Eq.~12
should be replaced
by a more complicated function of kinematical variables.

The differential cross-section is the most 
sensitive observable for detecting any kind of contact interactions
via the deviation from the SM.
But the parameter dependence of the angular distributions is quite involved.
To unravel it without calculating a lot of  angular distributions we
chose as more illuminative a set of integral characteristics.
They are: the relative deviation in the total cross-sections from the SM
values
\begin{equation}
\Delta(P_\mu)=\frac{\sigma(P_\mu)-\sigma_{SM}(P_\mu)}{\sigma_{SM}(P_\mu)},
\end{equation}
with $\sigma(P_\mu)$ being 
the polarized cross-section $\sigma(P_\mu)=(1-P_\mu)/2\,\cdot\sigma(\mu^-_L)+
(1+P_\mu)/2\,\cdot\sigma(\mu^-_R)$,
the forward-backward charge asymmetry
\begin{equation}
A_{FB}=\frac{\sigma_F-\sigma_B}{\sigma_F+\sigma_B},
\end{equation}
the left-right polarization asymmetry
\begin{equation}
A_{LR}=\frac{\sigma(\mu^-_L)-\sigma(\mu^-_R)}{\sigma(\mu^-_L)+\sigma(\mu^-_R)}
\end{equation}
and the polarized charge asymmetry
\begin{equation}
A^{FB}_{LR}=\frac{\sigma_F(\mu^-_L)+\sigma_B(\mu^-_R)-\sigma_F(\mu^-_R)-\sigma_B
(\mu^-_L)}{\sigma_F(\mu^-_L)+\sigma_B(\mu^-_R)+\sigma_F(\mu^-_R)+\sigma_B(\mu^-_L)}.
\end{equation}

We have calculated these observables (if not trivial)  for the processes 
$\mu^+\mu^-\to e^+e^-$, $\mu^+\mu^-$, $\tau^+
\tau^-$, $\bar bb$, $\bar cc$, $jet\,jet$ and $\mu^+\mu^-\to W^+W^-$, $ZH$,
$ZHH$ as functions of the parameter $\eta_1$ for the 
various values 
of ${\cal F}$. Under ``{\rm jets}'' we mean only these of the light and charmed
hadrons. Fig.\ 1 is a typical example of such a calculation for 
the process  $\mu^+\mu^-\to
e^+e^-$. Note that all the numerical results have been obtained using the 
exact Born expressions for differential cross-sections. Nevertheless
 Eqs.~10--12
give good approximations for both the qualitative and quantitative conclusions.
For all the processes (exept for $\mu^+\mu^-\to\mu^+\mu^-$ and the $W$ 
pair production) all 
the asymmetries have the similar behaviour. 
First of all, there exists a particular value
 of $\eta_1=\bar s^2/\bar c^2\simeq0.3$ when all the rescaling factors 
coincide  with each other
\begin{equation}
\kappa_1=\kappa_2=\kappa_3=1-\frac{\bar s^2}{\bar e^2}\frac{s}{{\cal F}^2},
\end{equation}
and thus all the asymmetries 
coincide with those
of the SM. The only way to unravel the contact interactions in this particular
case
is to study directly the total cross-sections. But there should be strong
natural reasons for this exceptional case to be realized.
Another particular value of $\eta_1=g_1^2{\cal F}^2/s$ provides the best 
case for 
studying the contact interactions, when all the asymmetries in all the processes
saturate their maximal values.

It is of no importance whether muon beam is polarized or not in the case of
fermion pair production. But it is not so for the processes $\mu^+\mu^-\to W^+W^-$,
$ZH$ and $ZHH$.
In  all the cases of bosons production one has
 $|\Delta(-1)|\ll|\Delta(+1)|$.
Hence 
one is lead to  conclude that it is preferable to work with the maximally 
right-handedly
polarized muons to observe as large deviations in the total cross-sections
from the SM as possible. Fig.\ 2 presents the deviations in the total 
cross-sections for the unpolarized muons, as well as for
 the right-handed muons with 
$P_\mu=0.8$. Here the Higgs boson mass is taken to be $m_H=200\,\,$GeV.
The results are quite insensitive to it for the light and intermidiate Higgs
boson.
One can see that the deviations for the right-handed polarizations are at 
least three times as large as these 
for the unpolarized muon beam.

To evaluate the statistical significance of the observed deviations consider,
e.g., the 
total cross-sections. Taking into account only statistical errors, 
let us introduce
the quantity $n_\sigma=\Delta N/\sqrt{N_{SM}}=(\Delta\sigma/\sigma_{SM})
\sqrt{\sigma_{SM}\int {\cal L}\,dt}$ that shows the number 
of the standard deviations
from the SM predictions. We take the integrated luminocity $\int {\cal L}\,dt$
expectedly to be 10$^3\,\,fb^{-1}$~\cite{atac}. 
Fig.~3 and Fig.~4 present the reach for the scale ${\cal F}$ at
$2\sigma$ statistical level (95\% C.L.) via the total cross-sections in the 
various channels. Note that the calculation for the $W^+W^-$ pair production 
has been made supposing the instrumental cut-off $-0.8\leq\cos\theta\leq 0.8$.
In the cases of both the $\mu^+\mu^-\to W^+W^-$ and $\mu^+\mu^-\to\mu^+\mu^-$ 
optimal values of cut-offs, equal to $-0.8\leq\cos\theta\leq 0.3$ and
$|\cos\theta|\leq 0.8$, respectively, have been chosen at the given 
instrumental ones.
 Here the reach is maximal due to the maximal
supression of the $t$ channel peak, at the statistics being 
still high enough. We see that the VBD can be tested for the unified
substructure scale ${\cal F}$ up to ${\cal O}(150\,\,$TeV)
in the processes $\mu^+\mu^-\to \bar ff$, up to 
${\cal O}(100\,\,$TeV) in the $\mu^+\mu^-$ annihilation into boson pairs and
up to ${\cal O}(40\,\,$TeV) in the process $\mu^+\mu^-\to ZHH$ (with 
 the right-handedly polarized muon beam). 
For comparison we present also the reach for the scale ${\cal F}$ at the 
$3\sigma$ statistical level (99\% C.L.). We see that it is not much lower, 
except for the channel $\mu^+\mu^-\to ZH$ with the  unpolarized muon 
beam~\footnote{It is to be studied to what extent the $P_\mu\neq 0$
effect could overwhelm an induced reduction in luminocity.}.

One can estimate the energy and luminocity dependence of the attainable scale
${\cal F}$ by equating the statistical uncertainty in the event number $\Delta
N\sim\sqrt{\int{\cal L}\,dt/s}$ and the expected number of additional 
events due to the contact interactions $\Delta N\sim\int{\cal L}\,
dt/{\cal F}^2$. This gives ${\cal F}\sim\sqrt[4]{s\int{\cal L}\,dt}$. 
Thus the decrease
of energy down to 2~TeV at the fixed luminocity would result in $\sqrt2$
decrease in the attainable scale ${\cal F}$. Hence 2~TeV collider is able 
to unravel the Deca-TeV substructure, too.

\paragraph{Anomalous Triple Gauge Interactions }

In addition to the  VBD interactions, a lot of other ``low energy'' residual
interactions is allowed in the scheme of the unified compositeness.
 In particular, the exotic triple gauge interactions (TGI)~\cite
{triple} are conceivable too and can contribute to the
$W^+W^-$ pair production. The question arises as to what extent the two
types of new interactions could imitate each other.
 
The anomalous TGI should  originate from a kind of the SM extension.
Here, the  SM symmetry $SU(2)_L\times U(1)_Y$
could be realized either linearly or nonlinearly.
In the case of the nonlinear realization (being still linear on the 
unbroken $U(1)_{em}$
subgroup), the nonlinearity scale $\Lambda$ is just the SM v.e.v.\ $v$.
Thus this kind of extension, in general, has nothing 
to do with the unified compositeness
we consider.
On the other hand, for the linear SM symmetry realization the scale 
$\Lambda$ is not
directly related with $v$ and could be as high as desired. Thus we chose 
it to be the
unified compositeness  scale ${\cal F}={\cal O}(10\,\,$TeV).

All the conceivable linearly realized residual
interactions are described by the $SU(2)_L\times U(1)_Y$ invariant operators
 built of the SM fields~\cite{leung,rujula}.
 All the operators
which are relevant to the anomalous TGI vertices are naturally expected
to be  ${\cal O}(g)$ or less in the gauge couplings, but one exception
 ${\cal O}_{WS}$. 
The latter stems from the nonlinear generalization of the field strengths in
the NMSM.
 The similar gauge kinetic terms of the
isotriplet $W$ and isosinglet $S$ bosons have no  gauge
couplings.
So the same must naturally happen for ${\cal O}_{WS}$, for its origin is
of the same nature.

Thus we have retained the ${\cal O}_{WS}$ operator alone and have chosen
 the proper
effective Lagrangian to be
\begin{equation}
{\cal L}_{eff}=\frac{C}{2}\frac{1}{{\cal F}^2}{\cal O}_{WS}
\equiv\frac{C}{2}\frac{1}{{\cal F}^2}
\phi^\dagger\frac{\tau_i}{2}\phi
W^i_{\mu\nu}S_{\mu\nu},
\end{equation}
where $C={\cal O}(1)$.
With account for  all the contributions from this operator we have found
 that the
deviations from the SM predictions even in this most enhanced
TGI case are much smaller then these in the VBD case.   
So the VBD is in fact dominant.

\section*{Conclusions}
The main results of our study are as follows:
\begin{itemize}

\item VBD of the SM gauge interactions is expected to be the universal
dominant low energy feature of the unified compositeness of leptons, quarks
and Higgs bosons.

\item VBD of the SM electroweak interactions can be tested at the 
4~TeV $\mu^+\mu^-$   collider for the unified
compositeness scale ${\cal F}$ up to ${\cal O}(150\,\,$TeV) in $\mu^+\mu^-\to
\bar ff$, up to ${\cal O}(100\,\,$TeV) in $\mu^+\mu^-\to ZH$, $W^+W^-$
and up to ${\cal O}(40\,\,$TeV) in the process $\mu^+\mu^-\to ZHH$. 

\item Processes $\mu^+\mu^-\to \bar ff$ with various final fermions and $\mu^+
\mu^-\to W^+W^-$, $ZH$, $ZHH$ are mutually complimentary. I.e., at any values of
compositeness scale ${\cal F}$ and parameter $\eta_1$ (but for $\eta_1
\simeq0.3$) one can choose the environments where the deviations from the SM
are not zero. More than that, these deviations are tightly correlated.

\item For $\mu^+\mu^-\to W^+W^-$, $ZH$ and $ZHH$ it is of importance 
to operate with 
the right-handed muons to observe as large deviations in the total
cross-sections as possible.

\item For $\mu^+\mu^-\to \mu^+\mu^-$ and $W^+W^-$ there exist the optimal angular
 cut-offs
 $|\cos\theta|\leq0.8$ and $-0.8\leq\cos\theta\leq0.3$, respectively, 
at which the attainable compositeness scale ${\cal F}$ is maximal.
\end{itemize}
 
We conclude that the future $\mu^+\mu^-$  collider
with the  total energy 4~TeV and the integrated luminocity 
10$^3\,\,fb^{-1}$ could present the definite answer 
about the existence of the Deca-TeV unified compositeness, or v.v.

\paragraph{Acknowledgement}
This work is supported in part by the Russian Foundation
for Basic Research (project No.~96-02-18122) and in part by 
the Competition Center for
Fundamental Natural Sciences (project No.~95-0-6.4-21).
One of us (Yu.F.P.) is
grateful for hospitality to Ankara University where this report was completed.

\paragraph{Note added}
After the report was submitted, we became aware of ref.~\cite{tikhonin} where
the process $\mu^+\mu^-\to ZH$ was studied in the framework of the SM.

\newpage
\section*{\bf Figure Captions}

\paragraph{Fig.~1} Process $\mu^+\mu^-\to e^+e^-$:
(a) relative deviations in the total unpolarized
cross-section marked with the values of the compositeness scale ${\cal F}$
in TeV; (b) forward-backward asymmetry; (c) left-right asymmetry.

\paragraph{Fig.~2}Relative deviations in the total cross-sections: (a) process
$\mu^+\mu^-\to W^+W^-$ with the $\mu^-$ polarization $P_\mu=0$;
(b) the same process with $P_\mu=0.8$;
(c) processes $\mu^+\mu^-\to ZH$ and $ZHH$  with $P_\mu=0$; (d) the same
processes with $P_\mu=0.8$.

\paragraph{Fig.~3} (a)~The reach at 95\% C.L.\ ($2\sigma$ statistical level)
for the compositeness scale ${\cal F}$,
vs.~the parameter $\eta_1$, via studying the total cross-sections of the
processes $\mu^+\mu^-\to \bar ff$  with $P_\mu=0$, where $f=e$, $\mu$,
$\tau$, $u$, $d$, $c$, $b$; (b) the same at the 99\% C.L.\ ($3\sigma$
statistical level).

\paragraph{Fig.~4} (a)~The same as in Fig.~3~(a) for the processes
$\mu^+\mu^-\to ZH$, $W^+W^-$ and $ZHH$ with two various polarizations $P_\mu$;
(b)~the same at the 99\% C.L.\ ($3\sigma$ statistical level).


\begin{thebibliography}{99}

\bibitem{unified}
V.~V.~Kabachenko and Yu.~F.~Pirogov, submitted to the 28th Int.\ Conf.\ on 
High Energy Physics, Warsaw, 1996; hep-ph/9612275, to be published
in {\it Yad.\ Fiz.}
\vspace{-2.5mm}

\bibitem{pirogov1}
Yu.~F.~Pirogov,  {\it Int.\ J.\ Mod.\ Phys.} {\bf A7}, 6473 (1992);
in  {\it Proc.\  of  the Int. Conf. ``Quarks~'92''}, Zvenigorod, 1992,
eds.\ D.~Yu.~Grigoriev {\it et al.} (World Scientific, Singapore, 1993)\ 375.
\vspace{-2.5mm}

\bibitem{pirogov2}
Yu.\ F.~Pirogov, {\it Mod.\ Phys.\ Lett.} {\bf A8}, 3129 (1993);
in {\it Proc.\ of the Int.\ School ``Particles and Cosmology''},
Baksan Valley, 1993, eds.\ E.~N.~Alexeev {\it et al.} (World Scientific,
 Singapore, 1994) p.\ 151.
\vspace{-2.5mm}

\bibitem{mu94}
{\it Proc.\ of the First Workshop on the Physics Potential and Development
of $\mu^+\mu^-$ Collider}, Napa, 1992, {\it Nucl.\ Instr.\ Meth.} {\bf A350},
24 (1994).
\vspace{-2.5mm}

\bibitem{bar95}
V.\ Barger {\it et al.}, MADPH-95-873 (1995), in {\it Proc.\ of the 
Second Workshop on the Physics Potential and Development
of $\mu^+\mu^-$ Collider}, Sausalito, 1994.
\vspace{-2.5mm}

\bibitem{gun95}
J.\ F.\ Gunion, UCD-95-35 (1995), in {\it Proc.\ of the Intern.\
Europhysics Conf.\ on High Energy Physics}, Brussels, 1995.
\vspace{-2.5mm}

\bibitem{bar97}
V.\ Barger, M.\ S.\ Berger, J.\ F.\ Gunion and T.\ Han, hep-ph/9704290,
in {\it Proc.\ of the 
Third Workshop on the Physics Potential and Development
of $\mu^+\mu^-$ Collider}, Santa Barbara, 1996.  
\vspace{-2.5mm}

\bibitem{coleman}
S.~Coleman, J.~Wess and B.~Zumino, {\it Phys.~Rev.} {\bf 177}, 2239 (1969);
C.~G.~Callan, S.~Coleman, J.~Wess and B.~Zumino, {\it ibid}, 2247.
\vspace{-2.5mm}

\bibitem{kugo}
M.~Bando, T.~Kugo, S.~Uehara, K.~Yamawaki  and  T.~Yanagida,
{\it Phys.\ Rev.\ Lett.} {\bf 54}, 1215 (1985); M.~Bando,  T.~Kugo  and
K.~Yamawaki, {\it Nucl.\ Phys.} {\bf B259}, 493 (1985); {\it Prog.\ 
Theor.\ Phys.} {\bf 73}, 1541 (1985);
{\it Phys.\ Rep.} {\bf 164}, 217  (1988).
\vspace{-2.5mm}

\bibitem{dada}
A.~D'Adda, P.~di Vecchia and M.~ L\"uscher, {\it Nucl.\ Phys.} {\bf B146}, 
63 (1978);
{\it Nucl.\ Phys.} {\bf B152}, 125 (1979);
A.~M.~Polyakov, Gauge Fields and Strings, {\it Contemporary Concepts in 
Physics}, v.~3 (Harwood Academic  Publishers,
1987) p.~139.
\vspace{-2.5mm}

\bibitem{kabach1}
V.~V.~Kabachenko and Yu.~F.~Pirogov,  {\it Int.\ J.\ Mod.\ Phys.}
{\bf A10}, 3187 (1995).
\vspace{-2.5mm}


\bibitem{kabach2}
V.~V.~Kabachenko and Yu.~F.~Pirogov, {\it Int.\ J.\ Mod.\ Phys.} {\bf A11},
2293 (1996).
\vspace{-2.5mm}

\bibitem{guryev}
D.~Guryev, V.~A.~Ilyin, V.~V.~Kabachenko and Yu.~F.~Pirogov, in preparation.
\vspace{-2.5mm} 

\bibitem{comphep}
 E.~E.~Boos, M.~N.~Dubinin, V.~A.~Ilyin, A.~E.~Pukhov and V.~I.~Savrin,
preprint SNUTP-94-116, hep-ph/9503280; E.~E.~Boos {\it et al.}, 
{\it Int.\ J.\ Mod.\ Phys.} {\bf C5} 615 (1994).

\vspace{-2.5mm}
\bibitem{atac}
M.~Atac, report at this Workshop.

\vspace{-2.5mm}
\bibitem{triple}
K.~Hagiwara, R.~D.~Peccei, D.~Zeppenfeld and K.~Hikasa, {\it Nucl. Phys.}
{\bf B282}, 253 (1987); M.~A.~Samuel {\it et al.}, {\it Phys.\ Rev.\ Lett.}
{\bf 67}, 9 (1991).
\vspace{-2.5mm}

\bibitem{leung}
C.~N.~Leung, S.~L.~Love and S.~Z.~Rao, {\it Z.~Phys.} {\bf C} --
{\it Part.\ and Fields} {\bf 31}, 433 (1986);
W.~Buchmuller and D.~Wyler, {\it Nucl.\ Phys.} {\bf B236}, 621 (1986).
\vspace{-2.5mm}

\bibitem{rujula}
A.~de~Rujula, M.~B.~Gavela, P.~Hernandes and E.~Masso, {\it Nucl. Phys.}
{\bf B384}, 3 (1992).

\vspace*{-2.5mm}
\bibitem{tikhonin}
V.\ A.\ Litvin and F.\ F.\ Tikhonin, IHEP 97-24 (1997).

\end{thebibliography}
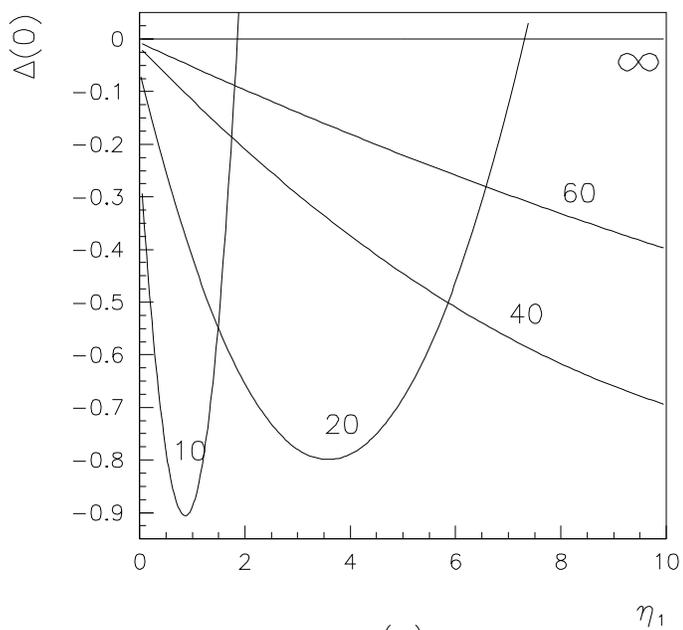
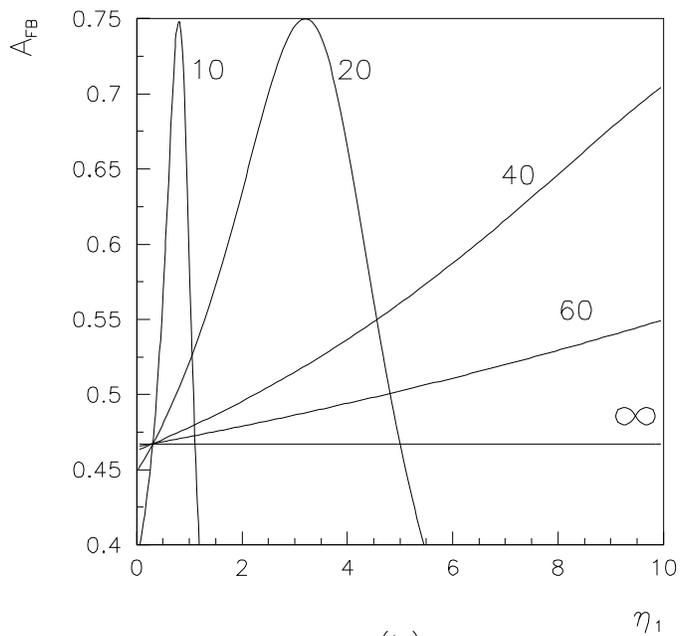
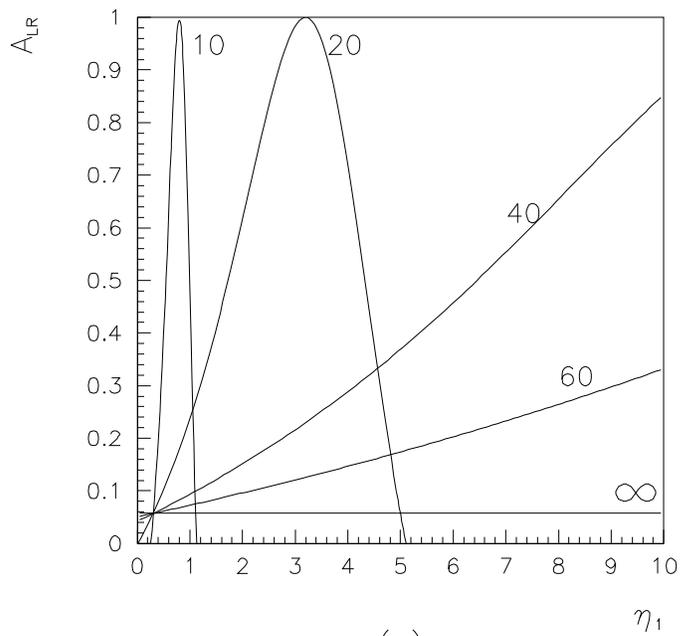
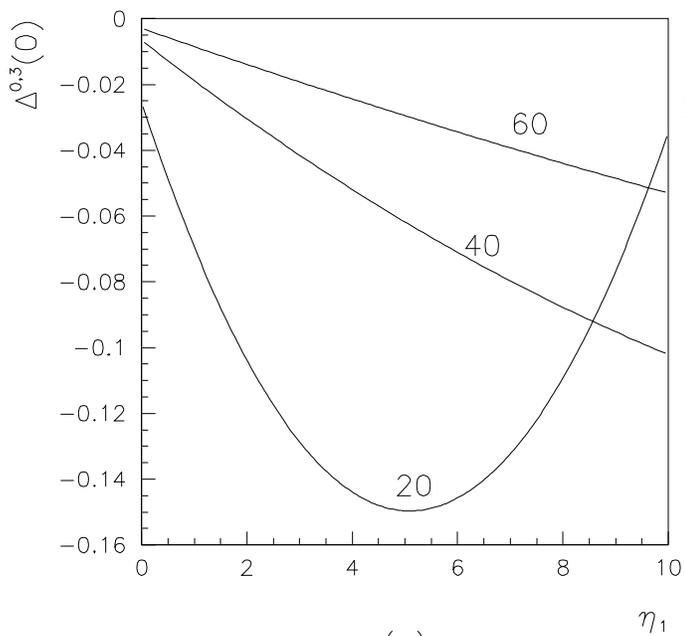
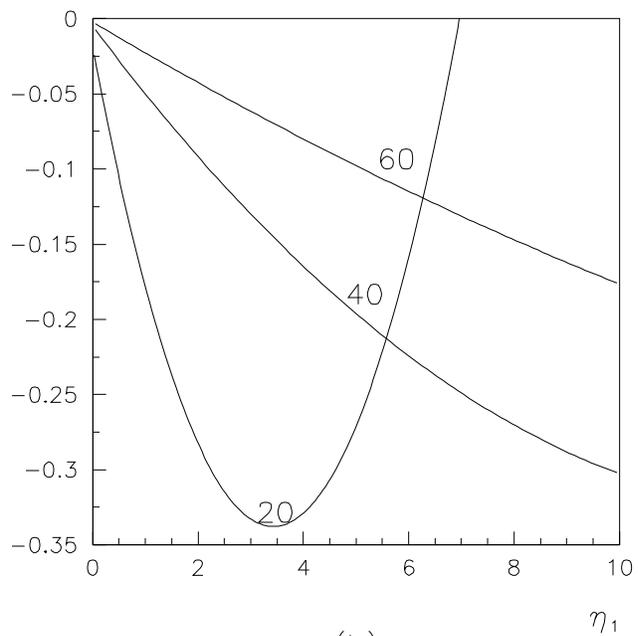
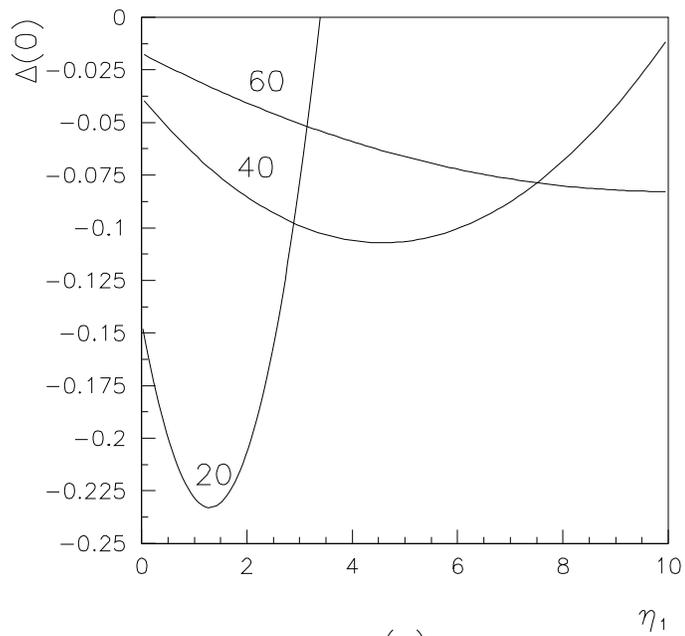
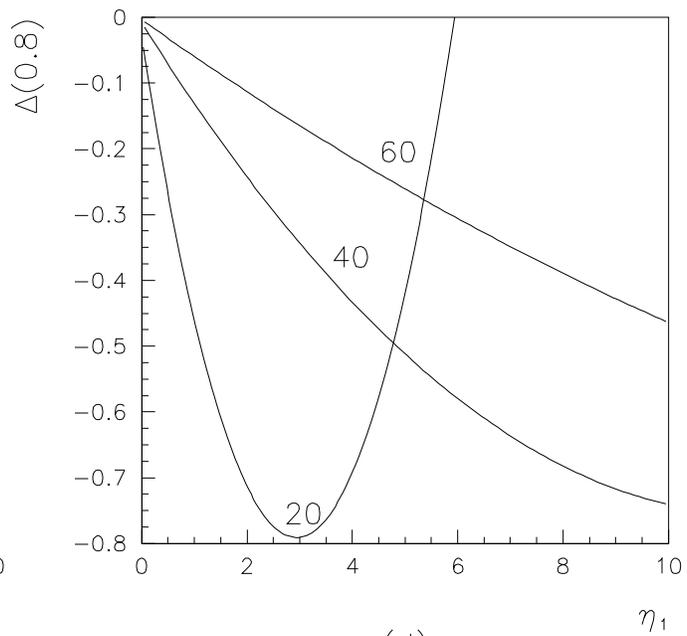
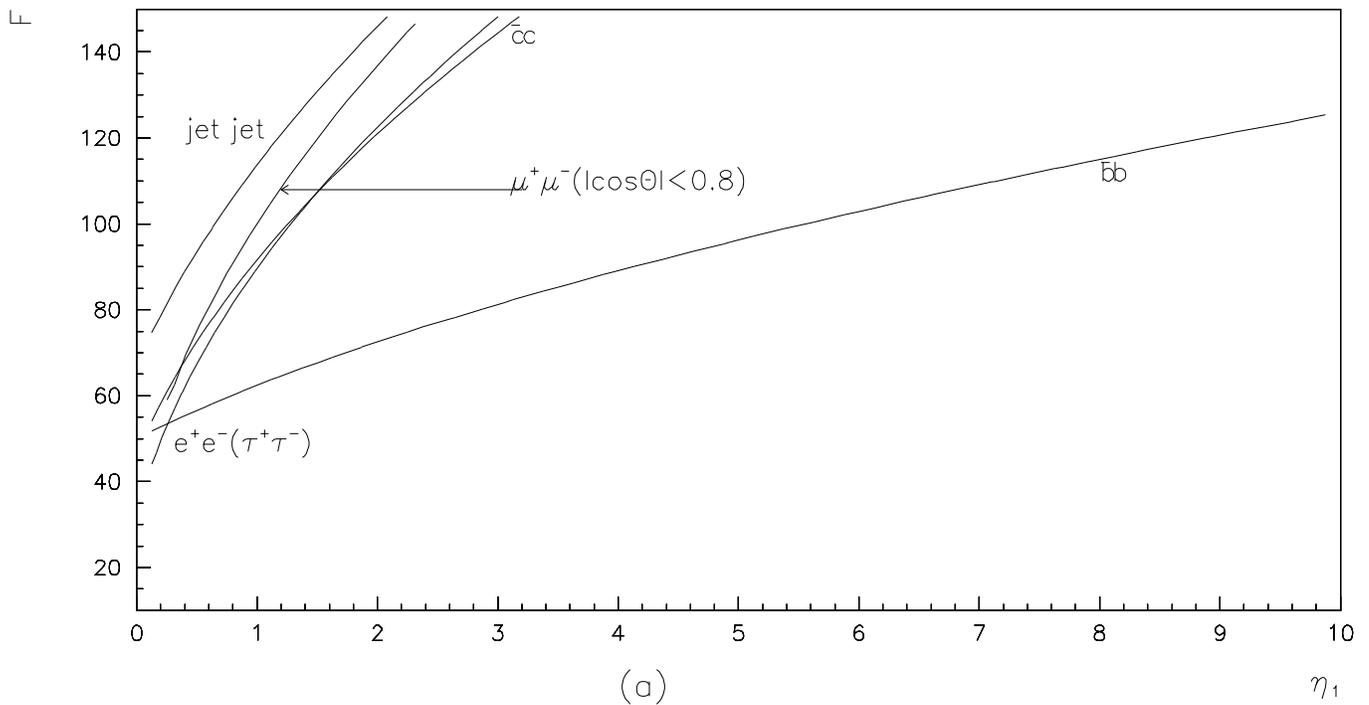
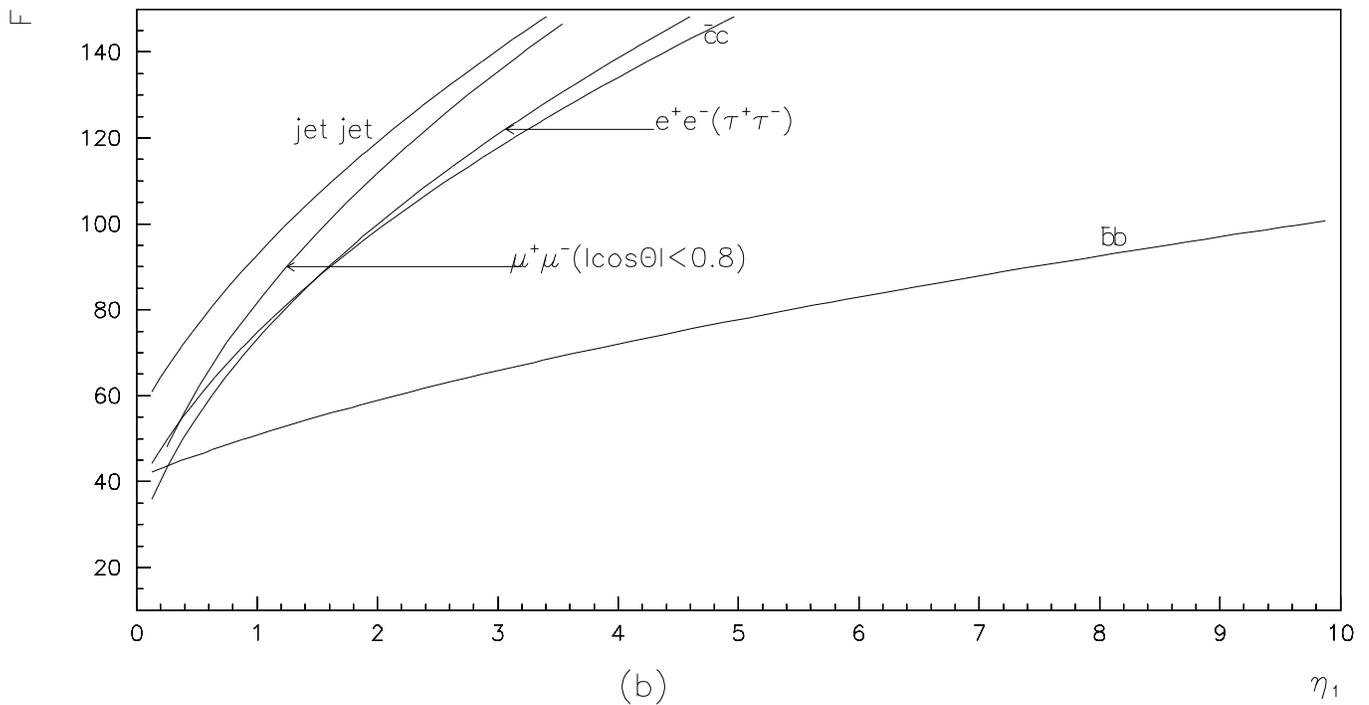
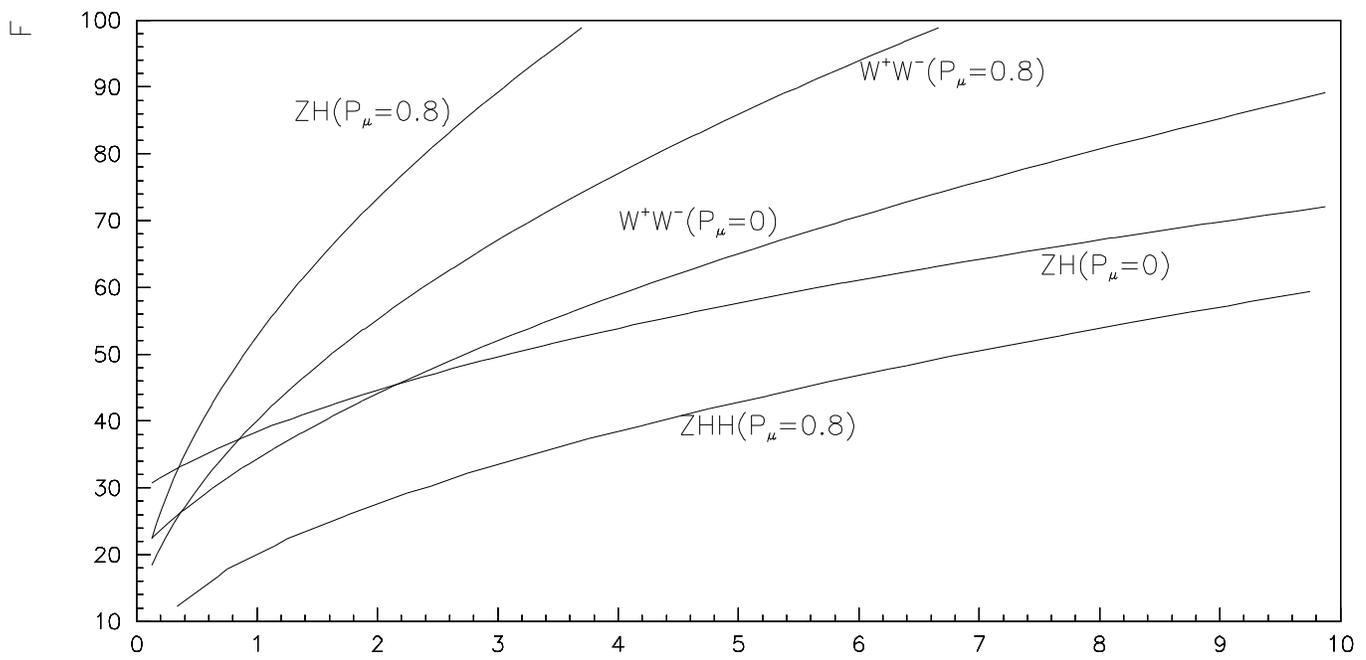
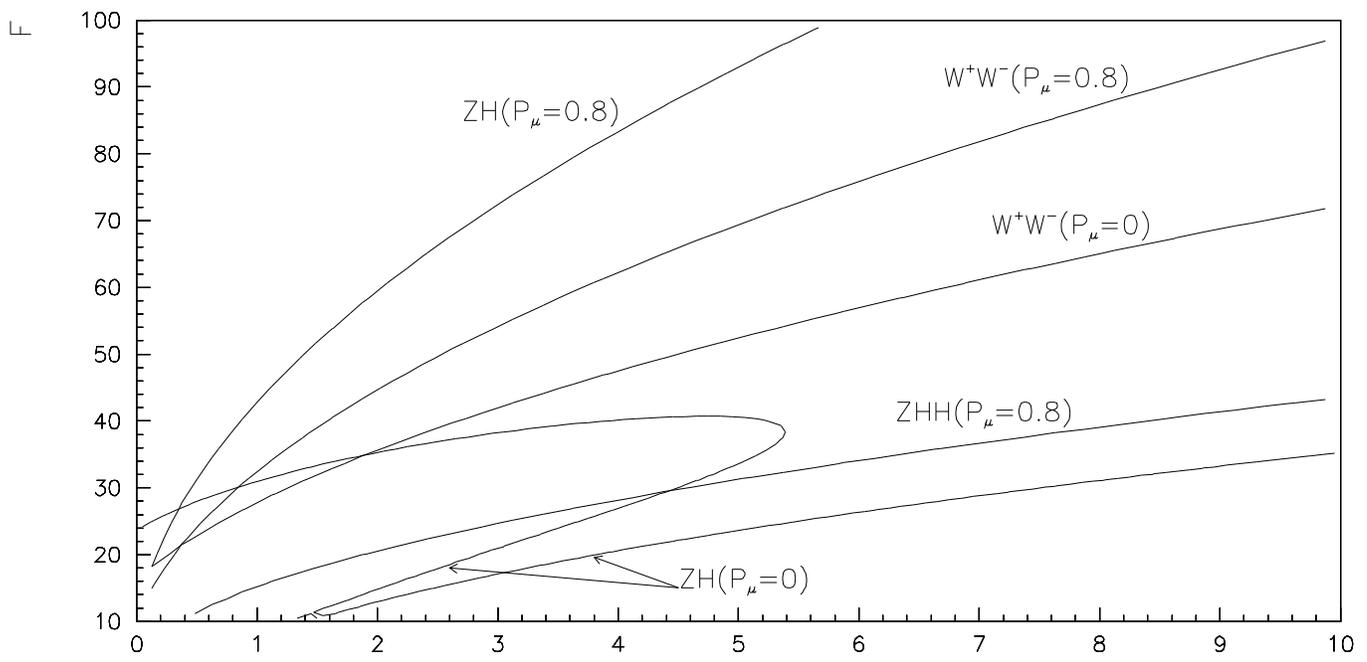
\end{document}